\documentclass[preprint,preprintnumbers, prd, floatfix, superscriptaddress,nofootinbib,utf8] {revtex4-1}
\usepackage{epsfig}
\usepackage{subfigure}
\usepackage{dcolumn}
\usepackage{bm}
\usepackage[usenames ,dvipsnames]{xcolor}
\usepackage{slashed}
\usepackage{graphicx,color}
\usepackage[bookmarksnumbered, pdfstartview=FitH,colorlinks,urlcolor=blue, citecolor=blue,linkcolor=blue] {hyperref}
\usepackage{appendix}
\usepackage{inputenc}
\usepackage{amsmath}
\begin{document}
\title{Power-Law Inflation Survives Observational Constraints}
\author{Yao Yu}
\email{Corresponding author: yuyao@cqupt.edu.cn}
\affiliation{Chongqing University of Posts \& Telecommunications, Chongqing, 400065, China}
 \affiliation{Department of Physics and Chongqing Key Laboratory for Strongly Coupled Physics, Chongqing University, Chongqing 401331, People's Republic of China}
\author{Wen-Zhang Feng}
\affiliation{Chongqing University of Posts \& Telecommunications, Chongqing, 400065, China}
\author{Hong-Song Xie}
\affiliation{Chongqing University of Posts \& Telecommunications, Chongqing, 400065, China}
\author{Han Zhang}
\affiliation{School of Physics, Zhengzhou University, Zhengzhou, Henan 450001, China}
\author{Bai-Cian Ke}
\email{Corresponding author: baiciank@ihep.ac.cn}
\affiliation{School of Physics, Zhengzhou University, Zhengzhou, Henan 450001, China}
\begin{abstract}
  Power-law inflation has stood as a classical model in inflationary cosmology
  since the early 1980s, prized for its exact analytical solutions and ability
  to naturally resolve the Big Bang theory's horizon and flatness problems
  through exponential expansion. However, its simplest form appears
  incompatible with modern precision observations, motivating increasingly
  complex alternatives. In this work, we demonstrate how previous predictions with
  power-law inflation considered only a particular solution of the field
  equations, and derive the complete set of general analytical solutions that
  satisfy current theoretical and observational constraints. This finding
  revitalizes power-law inflation as a viable framework, offering new
  possibilities for cosmological model-building while preserving its original
  mathematical elegance.
\end{abstract}
\maketitle
Inflation is the period of exponential expansion in the early Universe and
decides the initial conditions for the subsequent hot Big  Bang
era~\cite{Guth:1980zm}. Inflation explains why the universe appears spatially
flat, homogeneous, and isotropic on large scales, features that are difficult
to account for within the Big Bang theory
alone~\cite{Starobinsky:1980te,Linde:1983gd}. Moreover, inflation resolves the
horizon problem by allowing regions currently far apart to have once been in
causal contact, thus explaining the uniform temperature observed in the cosmic
microwave background~(CMB). It can also account for the origin of the
cosmological fluctuations and predicts a nearly scale-invariant power spectrum
for the fluctuations~\cite{Starobinsky:1979ty,Mukhanov:1981xt}.

Numerous different scenarios for inflation have been developed to describe
observational data, particularly the scalar spectral index $n_{\mathrm{s}}$ and
tensor-to-scalar ratio $r$ of primordial gravity waves. These scenarios
typically involve nonlinear ordinary differential equations, forcing most
relevant investigations to rely on numerical calculations with simplifying
approximations, such as the slow-roll approximation. A remarkable exception
was power-law inflation, developed by Lucchin and Matarrese in the early
1980s~\cite{er7,Lucchin:1984yf,Sahni:1988zb,Ratra:1989uv,Ratra:1989uz}, which
provided an elegant cosmological scenario with exact analytical solutions to
the inflationary field equations. Its exponential
potential~($V(\phi) \propto e^{-\lambda \phi}$) yielded power-law
expansion~($a(t)\propto t^p$) without requiring the slow-roll
approximation~\cite{Liddle:1994dx,Domenech:2023dxx}. This scenario's
mathematical simplicity enabled exact computation of primordial perturbations
while naturally solving the Big Bang theory's horizon and flatness problems
through prolonged expansion~($\geq 60$ e-foldings). Regrettably, this
theoretically elegant model has been effectively excluded by the latest CMB
measurements of $n_{\mathrm{s}}$ and $r$, diminishing its role in recent
inflationary studies and relegating it to a pedagogical tool with analytically
solvable solutions.

Previous calculations with power-law inflation, however, considered only a
particular solution, not the complete set of general solutions. This work
introduces a transform of the inflationary field equations into Abel's equation
of the first kind \cite{kamke1977,polyanin2003exact}. The exponential form of
the power-law potential renders this equation analytically solvable, allowing
us to obtain the general solutions. We demonstrate that these solutions can
satisfy both theoretical requirements and current observational constraints,
thereby restoring the viability of power-law inflation as a compelling
cosmological scenario.

In the flat Friedmann-Lema\^{\i}tre-Robertson-Walker metric, the canonical
homogeneous inflaton field $\phi$ obeys the Klein-Gordon equation of motion:
\begin{eqnarray}
  \ddot{\phi} + 3H\dot{\phi} + \frac{d V}{d \phi} &=& 0\,,\label{eq:EOM}
\end{eqnarray}
where overdots denote time $t$ derivatives, $V(\phi)$ is the inflationary
potential, $H\equiv \dot{a}/a$ is the Hubble parameter, and $a$ is the scale
factor. During inflation when the Universe is dominated by the inflaton field,
the Friedmann equation
\begin{eqnarray}
  H^2 &=& \frac{1}{3M_{\rm pl}^2} \left( \frac{1}{2} \dot{\phi}^2 + V(\phi) \right)\label{eq:Friedmann}
\end{eqnarray}
couples the Hubble parameter to the inflaton's kinetic and potential energy,
with $M_{\rm pl}\equiv\frac{1}{\sqrt{8\pi G}}$ the reduced Planck mass. The
dynamical equation for cosmic expansion is derived by combining
Eqs.~\eqref{eq:EOM} and~\eqref{eq:Friedmann} and using the identity
$\ddot{\phi} = \frac{d\dot{\phi}}{d\phi} \dot{\phi}$, written as
\begin{eqnarray}
  \frac{\mathrm{d}\dot{\phi}}{\mathrm{d}\phi} \dot{\phi} + 3 \sqrt{\frac{1}{3M_{\rm pl}^2} \left( \frac{1}{2} \dot{\phi}^2 + V(\phi) \right)} \dot{\phi} + \frac{\mathrm{d} V}{\mathrm{d} \phi} &=& 0\,.\label{eq:nonlinear}
\end{eqnarray}
The above equation constitutes a first-order nonlinear ordinary differential
equation for $\dot{\phi}(\phi)$. A strategic variable transformation could
convert the equation into a form better suited for analytical solutions.
This transformation proceeds through the following three steps:

\textbf{Step 1:} Equation~\eqref{eq:nonlinear} is rewritten as
\begin{equation}
\mathrm{d}\left[\sqrt{\frac{1}{2}\dot{\phi}^2 + V(\phi)}\right] + \frac{\sqrt{3}}{2M_{\text{pl}}}\dot{\phi}\mathrm{d}\phi = 0\,.\label{ab1}
\end{equation}

\textbf{Step 2:} Applying the substitution $V(\phi)=g^2(\phi)$, with
$\sqrt{\frac{1}{2}\dot{\phi}^2 + g^2(\phi)}=g(\phi)\cosh u(\phi)$ and
$\dot{\phi}=\sqrt{2}g(\phi)\sinh u(\phi)$, transforms Eq.~\eqref{ab1} into
  \begin{equation}
    \frac{\mathrm{d}u(\phi)}{\mathrm{d}\phi} + \frac{\mathrm{d}\ln g(\phi)}{\mathrm{d}\phi}\coth u(\phi) + \frac{\sqrt{6}}{2M_{\text{pl}}} = 0\,.\label{ab2}
  \end{equation}

\textbf{Step 3:} The final substitution
\begin{equation}
  y(\phi)=\coth u(\phi)=\frac{\sqrt{\dot{\phi}^2 + 2V(\phi)}}{\dot{\phi}}\,,(|y|>1)\label{eq:y_definition}
\end{equation}
yields
\begin{equation}
  \frac{\mathrm{d}y}{\mathrm{d}\phi} = (y^2-1)\left(\frac{y}{2}\frac{\mathrm{d}\ln V}{\mathrm{d}\phi} + \frac{\sqrt{6}}{2M_{\text{pl}}}\right)\,.\label{zd}
\end{equation}
This equation is known as the \textit{Abel equation of the first kind},
$dy/d\phi = f_3(\phi)y^3 + f_2(\phi)y^2 + f_1(\phi)y + f_0(\phi)$, where here
$f_3(\phi) = -f_1(\phi) = \tfrac{1}{2}\tfrac{d\ln V(\phi)}{d\phi}$ and
$f_2(\phi) = -f_0(\phi) = \tfrac{\sqrt{6}}{2M_{\text{pl}}}$. Similar
approaches have been utilized in Ref.~\cite{Muslimov:1990be,Yurov:2008sy}.


For the power-law inflation potential $V=V_0\exp(-\lambda\phi/M_{\text{pl}})$,
Eq.~\eqref{zd} can be solved via separation of variables.
It is convenient to
introduce the variable $\omega$ defined by $\omega = \sqrt{3}/y$. The general
exact solution is then given by
\begin{eqnarray}\label{yuy}
\left|\sqrt{3}-\omega\right|^{\frac{\lambda}{\lambda-\sqrt{6}}}\times\left|\sqrt{3}+\omega\right|^{\frac{\lambda}{\lambda+\sqrt{6}}}\times\left|\omega-\frac{\lambda}{\sqrt{2}}\right|^{\frac{2 \lambda^2}{6-\lambda^2}} &=& CV\,,
\end{eqnarray}
where $\lambda$ is a dimensionless parameter and $C \geq 0$ is an integration
constant. A similar result to the above expression was also obtained in
Ref.~\cite{2018msl} based on a different method.
  Note that $\omega^2$ corresponds to the first Hubble flow
  parameter~\cite{Hoffman:2000ue,Schwarz:2001vv}. Using Eqs.~\ref{eq:Friedmann}
  and~\ref{eq:nonlinear}, one can express it as
\begin{equation}
\omega^2 = -\frac{\dot{H}}{H^2} = 1 - \frac{a}{\dot{a}^2}\,\ddot{a},
\end{equation}
which confirms that $\omega^2$ coincides with the first Hubble flow parameter.
The inflationary condition $\ddot{a} > 0$ is therefore equivalent to
$\omega^2 < 1$, providing a clear physical interpretation of $\omega$ as a
parameter indicating whether the universe is undergoing inflation. The
previously studied exact solutions correspond to the special case $C = 0$ with
$\omega = \lambda / \sqrt{2}$, for which the first Hubble flow parameter is
constant. However, these $C=0$ solutions are ruled out by Planck 2018
data~($n_{\mathrm{s}} = 0.9649 \pm 0.0042$~\cite{Planck2018}) and a
combination of BICEP/Keck 2018 and Planck PR4
data~($r < 0.032$ at 95\% C.L.~\cite{Tristram:2021tvh, Tristram:2020wbi, BICEP:2021xfz}).

Our work therefore focuses on the case $C>0$ to derive general solutions that
are consistent with current observational constraints. The time derivative of
the scalar field can be obtained using Eq.~\eqref{eq:y_definition} as explicit
functions of $\omega$
\begin{eqnarray}\label{zy2}
  \dot{\phi} &=&\omega \sqrt{\frac{2V}{3-\omega^2}}\,.
\end{eqnarray}
Combining this with the Friedmann equation (Eq.~\eqref{eq:Friedmann}) yields
the Hubble parameter
\begin{eqnarray}\label{zy22}
H =\frac{1}{M_{\rm pl} \sqrt{\frac{V}{3-\omega^2}}}\,.
\end{eqnarray}
For $C > 0$, substitution of the general solution from Eq.~\eqref{yuy} gives
$\dot{\phi}$ and $H$ as
\begin{eqnarray}\label{zy3}
\dot{\phi} &=& \sqrt{\frac{2}{C}} \, \omega \,
\left|\sqrt{3}-\omega\right|^{\frac{\sqrt{6}}{2(\lambda-\sqrt{6})}}\times
\left|\sqrt{3}+\omega\right|^{-\frac{\sqrt{6}}{2(\lambda+\sqrt{6})}}\times
\left|\omega-\tfrac{\lambda}{\sqrt{2}}\right|^{\frac{\lambda^2}{6-\lambda^2}}\,, \nonumber\\
H &=& \frac{1}{\sqrt{C} M_{\text{pl}}}
\left|\sqrt{3}-\omega\right|^{\frac{\sqrt{6}}{2(\lambda-\sqrt{6})}}\times
\left|\sqrt{3}+\omega\right|^{-\frac{\sqrt{6}}{2(\lambda+\sqrt{6})}}\times
\left|\omega-\tfrac{\lambda}{\sqrt{2}}\right|^{\frac{\lambda^2}{6-\lambda^2}}\,.
\end{eqnarray}
Furthermore, Eq.~\eqref{zd} can be reformulated to express the evolution of
$\omega$ with the field
\begin{eqnarray}\label{zd2}
  \frac{\mathrm{d}\omega}{\mathrm{d}\phi} &=& -\frac{\sqrt{2}}{2M_{\rm pl}}\frac{1}{\omega}\left(3-\omega^2\right)\left(\omega-\frac{\lambda}{\sqrt{2}}\right)\,.
\end{eqnarray}
Combining the expression of $\dot{\phi}$ in Eq.~\eqref{zy3}, the time
derivative of $\omega$ is given by
\begin{eqnarray}\label{zy4}
  \dot{\omega}=\frac{\mathrm{d}\omega}{\mathrm{d}\phi}\dot{\phi} = \frac{1}{\sqrt{C}M_{\rm pl}}
  \left(\sqrt{3}-\omega\right)^{\frac{\sqrt{6}}{2}\frac{1}{\lambda-\sqrt{6}}+1}
  \left(\sqrt{3}+\omega\right)^{-\frac{\sqrt{6}}{2}\frac{1}{\lambda+\sqrt{6}}+1}
  \left(\frac{\lambda}{\sqrt{2}}-\omega\right)\left|\omega-\frac{\lambda}{\sqrt{2}}\right|^{\frac{\lambda^2}{6-\lambda^2}}\,.
\end{eqnarray}
Solving the above equation yields the explicit relation of $\omega$ and $t$
\begin{eqnarray}\label{eq:ful1}
  t&=&t_0+\frac{c_0}{M_{\rm pl}}\left|\frac{\lambda}{\sqrt{2}}-\omega\right|^{-\frac{\lambda^2}{6-\lambda^2}}\nonumber\\
  &&\times F_1\left(-\frac{\lambda^2}{6-\lambda^2};\, 1-\frac{1}{2}\frac{\sqrt{6}}{\sqrt{6}-\lambda}, 1-\frac{1}{2}\frac{\sqrt{6}}{\sqrt{6}+\lambda};\, 1-\frac{\lambda^2}{6-\lambda^2};\, \frac{\omega-\frac{\lambda}{\sqrt{2}}}{\sqrt{3}-\frac{\lambda}{\sqrt{2}}}, -\frac{\omega-\frac{\lambda}{\sqrt{2}}}{\sqrt{3}+\frac{\lambda}{\sqrt{2}}}\right)\,,
\end{eqnarray}
where
\begin{eqnarray}\label{eq:ful11}
c_0=\sqrt{C}M^2_{\rm pl}\frac{6-\lambda^2}{\lambda^2}\left(\sqrt{3}-\frac{\lambda}{\sqrt{2}}\right)^{-\frac{\sqrt{6}}{2}\frac{1}{\lambda-\sqrt{6}}-1}\left(\sqrt{3}+\frac{\lambda}{\sqrt{2}}\right)^{\frac{\sqrt{6}}{2}\frac{1}{\lambda+\sqrt{6}}-1}
\end{eqnarray}
and $F_1(a;\, b_1, b_2;\, c;\, x, y)$ denotes the Appell hypergeometric
function  of the first
kind~\cite{AskeyBateman:2020,Wikipedia:AppellSeries}, and $t_0$ is an
integration constant. Accordingly, the dynamical behavior of the field
$\phi(t)$ in time $t$ is derived in the parametric form of Eq.~\eqref{eq:ful1}
and the following equation, which is obtained from Eq.~\eqref{yuy}
\begin{eqnarray}\label{yuy2}
\phi &=\frac{M_{\rm pl}}{\lambda}\left[\frac{\lambda}{\sqrt{6}-\lambda}\ln\left|\sqrt{3}-\omega\right|-\frac{\lambda}{\sqrt{6}+\lambda}\ln\left|\sqrt{3}+\omega\right|-2\frac{\lambda^2}{6-\lambda^2}\ln\left|\frac{\lambda}{\sqrt{2}} - \omega\right|+\ln(CV_0)\right]\,.
\end{eqnarray}

For completeness, we also analyze the $C=0$ case to provide a comprehensive
theoretical comparison. In this limit, $\omega$ is a constant,
$\omega=\tfrac{\lambda}{\sqrt{2}}$, contrasting with the time-dependent
behavior found for $C>0$. Equations~\eqref{zy2} and \eqref{zy22} yield the
analytical form the time derivative of the scalar field and the Hubble
parameter as
\begin{eqnarray}\label{zz1}
\dot{\phi} &=& \lambda \sqrt{\frac{2V_0}{6-\lambda^2}}
\exp\!\left(-\frac{\lambda}{2} \frac{\phi}{M_{\text{pl}}}\right), \nonumber\\
H &=& \frac{1}{M_{\text{pl}}}
\sqrt{\frac{2V_0}{6-\lambda^2}}
\end{eqnarray}
Consequently, the dynamical behavior of the field $\phi(t)$ in time $t$ is
given by
\begin{eqnarray}\label{ui1}
  \phi &=& \frac{2M_{\rm pl}}{\lambda} \ln\left[\frac{\lambda^2}{2M_{\rm pl}}\sqrt{\frac{2V_0}{6-\lambda^2}}\right]+\frac{2M_{\rm pl}}{\lambda} \ln[t+t_1]\,,
\end{eqnarray}
where $t_1$ is an integration constant. The $C=0$ results cannot be obtained by
taking the $C\to 0$ limit of the $C>0$ solutions, as this leads to an
indeterminate $0/0$ form. This reveals an intrinsic analytic discontinuity; the
analytic expression valid for $C>0$ cannot be continuously extended to $C=0$.
Nevertheless, the $C=0$ case remains structurally connected to the $C>0$ case
and should be interpreted as its non-smooth counterpart rather than as an
entirely independent solution.

A complete understanding of the inflaton's behavior requires analysis of
several key features: stability conditions (if an attractor), the number of
e-foldings (typically $\gtrsim 60$ for successful inflation), super-Hubble
evolution of perturbations, and the observable spectra (the scalar spectral
index $n_{\mathrm{s}}$ and tensor-to-scalar ratio $r$). These features are now
examined. This work concentrates on $|\lambda|<\sqrt{6}$ for clarity, a range
that will later be shown to align with current observational constraints.
The stability conditions are analyzed by assuming $\phi_c$ is a stable fixed
point, a special type of an attractor, which satisfies
\begin{eqnarray}\label{yu5}
  \dot{\phi} \big|_{\phi=\phi_c}&=&0 ,\,\,\,\,\frac{d\dot{\phi}}{d\phi}\Big|_{\phi=\phi_c}<0\,.
\end{eqnarray}
The explicit expression for $\dot{\phi}$ in Eq.~\eqref{zy3} is used to identify
candidate fixed points. Their stability is then determined by evaluating the
sign of the derivative $\frac{d\dot{\phi}}{d\phi}$ at these points. The
derivative $\frac{d\dot{\phi}}{d\phi}$ is computed via the chain rule,
$\frac{\mathrm{d}\dot{\phi}}{\mathrm{d}\phi} =\frac{\mathrm{d}\dot{\phi}}{\mathrm{d}\omega}\frac{\mathrm{d}\omega}{\mathrm{d}\phi}$,
using Eq.~\eqref{zy3} for $\dot{\phi}$ and Eq.~\eqref{zd2} for
$\frac{\mathrm{d}\omega}{\mathrm{d}\phi}$. The explicit form of
$\frac{d\dot{\phi}}{d\phi}$ is derived to be
\begin{eqnarray}
  \frac{\mathrm{d}\dot{\phi}}{\mathrm{d}\phi} &=&\frac{\mathrm{d}\dot{\phi}}{\mathrm{d}\omega}\frac{\mathrm{d}\omega}{\mathrm{d}\phi}=\dot{\phi}\frac{\mathrm{d}\omega}{\mathrm{d}\phi}\left[\frac{1}{\omega}
    +6\frac{\frac{\lambda}{\sqrt{2}}+\omega}{\left(6-\lambda^2\right)\left(3-\omega^2\right)}+\frac{\lambda^2}{6-\lambda^2}\frac{1}{\omega-\frac{\lambda}{\sqrt{2}}}\right]\,.
\end{eqnarray}

\begin{figure}[t!]
\includegraphics[width=3in]{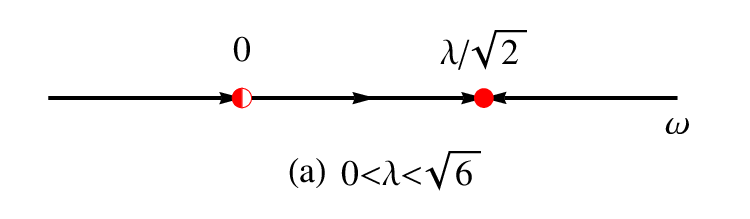}
\includegraphics[width=3in]{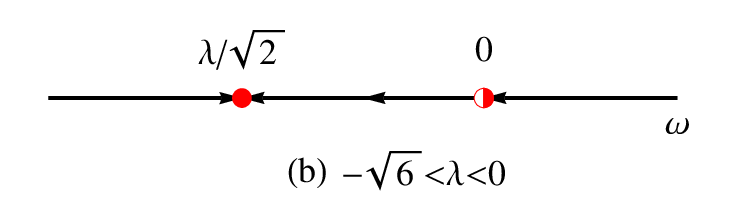}
\includegraphics[width=3in]{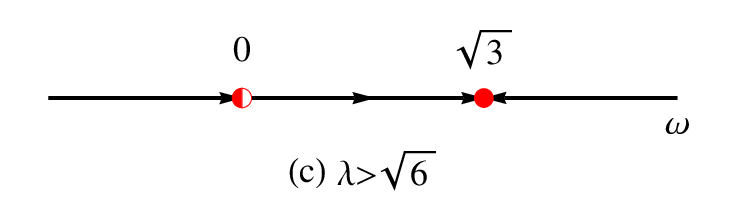}
\includegraphics[width=3in]{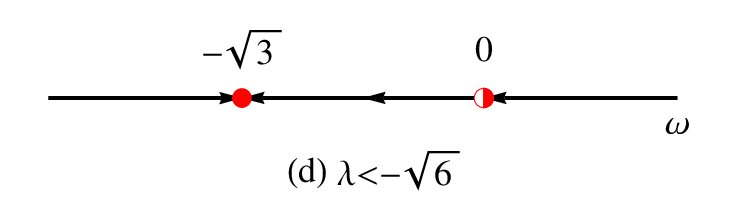}
\caption{Attractor behaviors for various values of $\lambda$}\label{tu2}
\end{figure}

Distinct attractor behaviors emerge for different $\lambda$ regimes,
illustrated in Fig.~\ref{tu2}:
\begin{itemize}
\item $0 < \lambda < \sqrt{6}$: When $\omega < 0$, $\omega \to 0$ (semi-stable fixed point), when $\omega > 0$, $\omega \to\lambda/\sqrt{2}$ (stable fixed point)
\item $-\sqrt{6} < \lambda < 0$:When $\omega > 0$, $\omega \to 0$ (semi-stable fixed point),when $\omega < 0$, $\omega \to\lambda/\sqrt{2}$ (stable fixed point)
\item $\lambda > \sqrt{6}$: When $\omega < 0$, $\omega \to 0$ (semi-stable fixed point); when $\omega > 0$, $\omega \to \sqrt{3}$ (stable fixed point).
\item $\lambda < -\sqrt{6}$: When $\omega > 0$, $\omega \to 0$ (semi-stable fixed point); when $\omega < 0$, $\omega \to -\sqrt{3}$ (stable fixed point).
\end{itemize}

The limiting behavior of $\frac{\mathrm{d}\dot{\phi}}{\mathrm{d}\phi}$ as
$\omega \to 0$ are considered from both $\omega > 0$ and $\omega < 0$
directions, since $\frac{\mathrm{d}\dot{\phi}}{\mathrm{d}\phi}$ exhibits a
singularity at $\omega = 0$. For instance, when $0 < \lambda < \sqrt{6}$, only
trajectories with $\omega > 0$ lies in the observationally allowed region; in
this physical regime, the semi-stable point behaves as \emph{repelling},
although mathematically it would become \emph{attracting} for $\omega < 0$, a
region not physically relevant. Once the value of $\omega$ is specified, the
corresponding value of $\phi(\omega)$ in Eqs.~\eqref{yuy} or \eqref{yuy2} can
be determined. For $0 < \lambda < \sqrt{6}$, as $\omega \to \lambda/\sqrt{2}$,
the potential $V$ approaches zero, corresponding to $\phi_c \to +\infty$. In
contrast, for $-\sqrt{6} < \lambda < 0$, $\phi_c \to -\infty$.

The e-folding number $N$ quantifies the exponential growth of the scale factor
$a(t)$ of the universe and the Hubble parameter $H=\dot{a}/a$ represents the
instantaneous expansion rate of the universe. For inflation from starting time
$t_i$ to ending $t_e$, $N$ for the $C>0$ case is given by
\begin{eqnarray}
  N &\equiv& \ln\frac{a(t_e)}{a(t_i)}=\int_{t_{i}}^{t_{e}}H dt=\int_{\omega_i}^{\omega_e}\frac{H}{\dot{\phi}} \frac{d\phi}{d\omega}d\omega=\int_{\omega_i}^{\omega_e}\frac{d\omega}{(\omega^2-3)\left(\omega-\frac{\lambda}{\sqrt{2}}\right)}\,.
\end{eqnarray}
Then one has
\begin{eqnarray}\label{er2}
  N &=& \frac{\sqrt{6}\lambda}{36-6\lambda^2}\ln\frac{\sqrt{3}-\omega}{\sqrt{3}+\omega}+\frac{1}{6-\lambda^2}\ln\frac{3-\omega^2}{\left(\omega-\frac{\lambda}{\sqrt{2}}\right)^{2}}\Big| ^{\omega_e}_{\omega_i}\,.
\end{eqnarray}

Consequently, the scale factor's evolution $a(t)$ follows the parametric form
of Eq.~\eqref{eq:ful1} with
\begin{equation}\label{ea1}
a =a_0\left|\sqrt{3}-\omega\right|^{\frac{\sqrt{6}}{6(\sqrt{6}-\lambda)}} \times\left|\sqrt{3}+\omega\right|^{\frac{\sqrt{6}}{6(\sqrt{6}+\lambda)}}\times\left|\omega-\frac{\lambda}{\sqrt{2}}\right|^{-\frac{2}{6-\lambda^2}}\,.
\end{equation}
As for $C=0$, the e-folding number is
\begin{eqnarray}
  N &\equiv& \ln\frac{a(t_e)}{a(t_i)}=\int_{t_{i}}^{t_{e}}H dt=\int_{\phi_i}^{\phi_e}\frac{H}{\dot{\phi}} d\phi=\frac{1}{\lambda M_{\rm pl}}(\phi_e-\phi_i)\,.
\end{eqnarray}
Combining this with Eq.~\eqref{ui1}, the scale factor is written as
\begin{eqnarray}\label{ea2}
a(t)&=&a_1(t+t_1)^{\frac{2}{\lambda^2}}\,,
\end{eqnarray}
where $a_1$ is an integration constant. According to Eqs.~\eqref{ea1}
and~\eqref{ea2}, this model does not naturally provide a graceful exit from
inflation, and an additional mechanism to end inflation must be introduced. It
is common in the literature to treat the inflationary phase and its
termination separately, because observational constraints mainly depend on the
dynamics during inflation rather than the specifics of its ending. For
instance, constant-roll inflation~\cite{Motohashi:2014ppa,Motohashi:2017aob}
likewise lack a built-in exit mechanism and require an additional scheme, such
as that described in Ref.~\cite{Linde:1993cn}, to terminate inflation.

\begin{figure}[t!]
\centering
\includegraphics[width=2.0in]{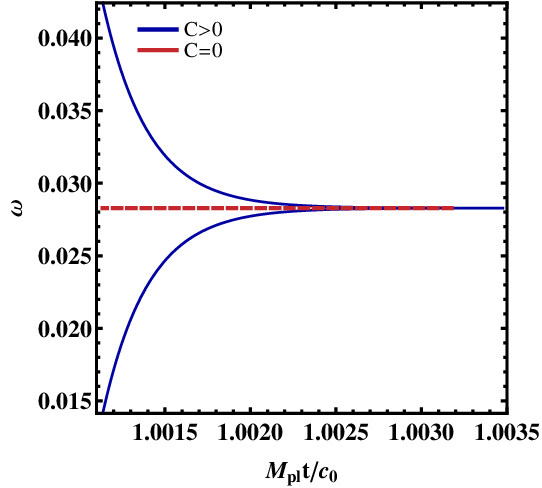}
\includegraphics[width=2.0in]{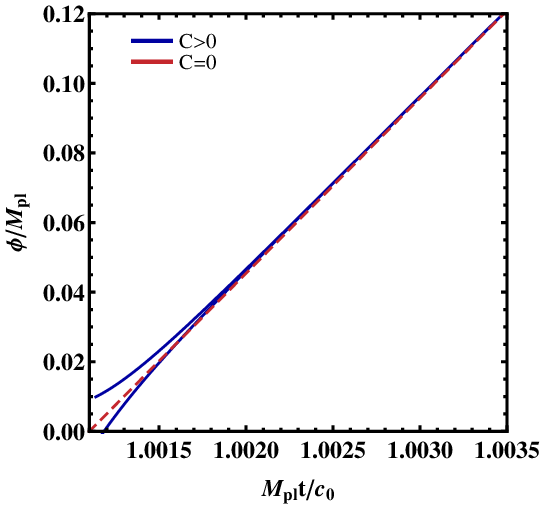}
\includegraphics[width=2.0in]{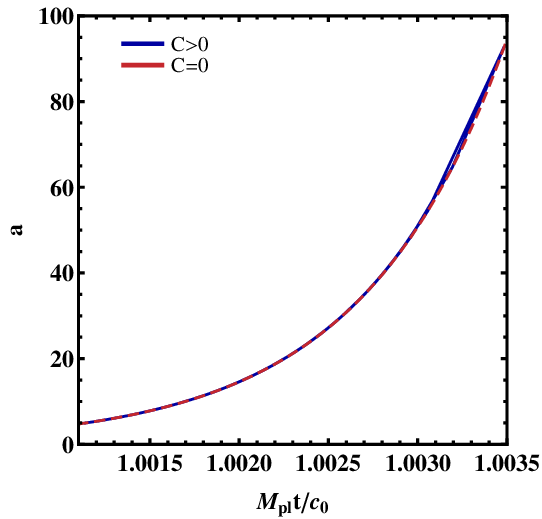}
\caption{
  Numerical solutions of $\omega(t)$, $\phi(t)$, and $a(t)$.  The blue curves
  correspond to the $C>0$ case, while the red curves represent $C=0$. The value
  of $\lambda$ is fixed to 0.04 because it leads to model parameters that
  satisfy observational constraints on $n_{\mathrm{s}}$ and $r$. Different
  $\lambda$ values would merely stretch or compress the curves. The horizontal
  axis $M_{\rm pl} t / c_0$ is used to removes rescaling caused by $C$ and keeps
  the time variable dimensionless. For the $C=0$ case we conventionally set
  $c_0 = 1$; for the $C>0$ case, $c_0$ is defined in Eq.~\ref{eq:ful11}, and 
  integration constants are chosen as $t_0 = 0$ and $a_0 = 1$, where $t_0$
  simply sets the horizontal origin without affecting the physical evolution.
}
\label{tu3}
\end{figure}

The analytical solutions for $\omega(t)$, $\phi(t)$, and $a(t)$ are presented
in Fig.~\ref{tu3}, which compares the $C>0$ and $C=0$ results. While the
corresponding evolutions are nearly identical across most of the inflationary
history, they diverge at the initial stage, following distinct trajectories
before converging. During the approach to the stable point, $\omega(t)$ tends
to a constant, $\phi(t)$ increases, and $a(t)$ exhibits convex growth,
signaling an accelerated expansion. As will be shown, the early-time
trajectories along which the solutions approach their stable points are
critical, leading to significantly different cosmological observables, such as
the scalar spectral index $n_{\mathrm{s}}$ and the tensor-to-scalar ratio $r$.

The comoving curvature perturbation $\zeta$ is the key variable that describes
how scalar field fluctuations influence the geometry of spacetime. Through
appropriate gauge choices and variable redefinitions, the system of
perturbations can be reduced to a single dynamical degree of freedom,
$\nu = z\zeta$ and $z = a\dot{\phi}/H$. The variable $\nu$ can be
reconstructed via the Fourier transform,
\begin{eqnarray}
\nu(\tau, \vec{x}) = \int \frac{d^3k}{(2\pi)^3} \, \nu_k(\tau) \, e^{i\vec{k} \cdot \vec{x}},
\end{eqnarray}
Here, $\tau\equiv \int a^{-1}dt$ is the conformal time and $v_k$ is the
canonically normalized, gauge-invariant scalar perturbation in Fourier space.
The evolution of $\nu_k$ is governed by the Mukhanov-Sasaki
equation~\cite{Mukhanov:1985rz,Sasaki:1986hm}
\begin{eqnarray}\label{eq:evolve_nuk}
  \nu_k'' +\left(k^2-\frac{z''}{z}\right)\nu_k= 0\,,
\end{eqnarray}
where a prime denotes the derivative with respect to $\tau$. The scalar
spectral index $n_{\mathrm{s}}$ characterizes the scale dependence of the
scalar power spectrum $\mathcal{P}_\mathcal{R}(k)$~\cite{Stewart:1993bc}. With
$\frac{z''}{z}\equiv \frac{\nu_R^2-1/4}{\tau^2}$, $n_{\mathrm{s}}$ can be
calculated by
\begin{eqnarray}\label{eq:ns}
n_{\mathrm{s}} - 1 \equiv \frac{d \ln \mathcal{P}_\mathcal{R}(k)}{d \ln k}=3-2\nu_R\,.
\end{eqnarray}
The validity of Eq.~\eqref{eq:ns} relies on the condition that the quantity
$\frac{z''}{z}\tau^2$ remains approximately constant with respect to $\tau$.
This approximation will be justified after Eq.~\eqref{eq:er6}.


In terms of the slow-roll parameters, the potential in the Mukhanov-Sasaki
equation reads~\cite{Hoffman:2000ue,Schwarz:2001vv}
\begin{eqnarray}\label{er3}
  \frac{z''}{z} &=& a^2H^2\left(2+\epsilon_1+\frac{3}{2}\epsilon_2+\frac{1}{4}\epsilon_2^2-\frac{1}{2}\epsilon_1\epsilon_2+\frac{1}{2}\epsilon_2\epsilon_3\right)\,,
\end{eqnarray}
with
\begin{eqnarray}
  \epsilon_1 &\equiv&-\frac{\dot{H}}{H^2}= \omega^2 ,\,\,\,\,\epsilon_2 \equiv \frac{\dot{\epsilon}_1}{H\epsilon_1}=-\frac{2}{\omega}\left(3-\omega^2\right)\left(\omega-\frac{\lambda}{\sqrt{2}}\right)\,, \\
  \epsilon_3 &\equiv&\frac{\dot{\epsilon}_2}{H\epsilon_2}= 4\omega^2-6+\frac{3\lambda}{\sqrt{2}\omega}\left(1-\omega^2\right)\,,
\end{eqnarray}
and the conformal time reads
\begin{eqnarray}\label{er5}
  \tau &=& \int^{t}_{\infty} \frac{1}{a }dt=\int^{\omega}_{\frac{\lambda}{\sqrt{2}}} \frac{1}{a\dot{\phi}\frac{d\omega}{d\phi} }d\omega\nonumber\\
  &=&\pm\frac{M_{\rm pl}\sqrt{C}}{a_0}\int^{\omega}_{\frac{\lambda}{\sqrt{2}}}\left|\sqrt{3}-\omega\right|^{\frac{\sqrt{6}}{3(\sqrt{6}-\lambda)}-1}\times \left|\sqrt{3}+\omega\right|^{\frac{\sqrt{6}}{3(\sqrt{6}+\lambda)}-1}\times\left|\omega-\frac{\lambda}{\sqrt{2}}\right|^{-\frac{4}{6-\lambda^2}}d\omega\,.
\end{eqnarray}
Combining Eqs.~\eqref{zy3},~\eqref{ea1},~\eqref{er3}, and~\eqref{er5},
one can find
\begin{eqnarray}\label{eq:er6}
  \frac{z''}{z}\tau^2 &=& \frac{4}{(\lambda^2-2)^{2}\omega^2}\left(\frac{\sqrt{3}-\frac{\lambda}{\sqrt{2}}}{\sqrt{3}-\omega}\right)^{\frac{2}{3}\frac{\sqrt{6}}{\sqrt{6}-\lambda}}
  \left(\frac{\sqrt{3}+\frac{\lambda}{\sqrt{2}}}{\sqrt{3}+\omega}\right)^{\frac{2}{3}\frac{\sqrt{6}}{\sqrt{6}+\lambda}}\nonumber\\
  &\times&\left[2\left(\omega-\frac{\lambda}{\sqrt{2}}\right)^2\left(9-9\omega^2+2\omega^4\right)+\omega^2\left(2+\omega^2\right)\right]\nonumber\\
  &\times& F_1\left(\frac{2-\lambda^2}{6-\lambda^2};\, 1-\frac{1}{3}\frac{\sqrt{6}}{\sqrt{6}-\lambda}, 1-\frac{1}{3}\frac{\sqrt{6}}{\sqrt{6}+\lambda};\, 1+\frac{2-\lambda^2}{6-\lambda^2};\, \frac{\omega-\frac{\lambda}{\sqrt{2}}}{\sqrt{3}-\frac{\lambda}{\sqrt{2}}}, -\frac{\omega-\frac{\lambda}{\sqrt{2}}}{\sqrt{3}+\frac{\lambda}{\sqrt{2}}}\right)^2\,.
\end{eqnarray}
Then, the scalar spectral index $n_{\mathrm{s}}$ can be determined with
Eqs.~\eqref{eq:ns} and~\eqref{eq:er6}.

The quantity $\dfrac{z''}{z}\,\tau^2$ must be approximately constant in $\tau$
for the slow-roll-like approximation in Eq.~\eqref{eq:ns} to be valid. Since
$\dfrac{z''}{z}\,\tau^2$ is a function of $\omega$, we analyze its behavior by
examining $d\omega/d\tau$ near the stable point $\omega=\lambda/\sqrt{2}$.
Following Eq.~\eqref{er5}, one has
\begin{equation}\label{domega_dtau}
  \frac{d\omega}{d\tau} = \left(\frac{d\tau}{d\omega}\right)^{-1}
    = \pm \frac{a_0 }{M_{\rm pl}\sqrt{C}}
      \left|\sqrt{3}-\omega\right|^{-\frac{\sqrt{6}}{3(\sqrt{6}-\lambda)}+1}
      \left|\sqrt{3}+\omega\right|^{-\frac{\sqrt{6}}{3(\sqrt{6}+\lambda)}+1}
      \left|\omega-\frac{\lambda}{\sqrt{2}}\right|^{\frac{4}{6-\lambda^2}} \,.
\end{equation}
Near $\omega\simeq\lambda/\sqrt{2}$ with $|\lambda|<\sqrt{6}$, the above
equation implies $d\omega/d\tau\simeq0$. Thus, $\omega$ remains nearly constant
without singular behavior. The region $|\lambda| > \sqrt{6}$ is excluded, as it
yields values of $n_{\mathrm{s}}$ and $r$ inconsistent with observational
constraints.

The evolution of tensor perturbations (or primordial gravity waves) can
likewise be formulated~\cite{Mukhanov:1990me}. Each Fourier mode of the tensor
perturbation, with amplitude $u_k$, obeys
\begin{eqnarray}
  u_k'' +\left(k^2-\frac{a''}{a}\right)u_k&=& 0\,.
\end{eqnarray}
With $\frac{a''}{a}\equiv \frac{\nu_{T}^2-1/4}{\tau^2}$, the tensor-to-scalar
ratio $r$ is given as~\cite{Lidsey:1995np}
\begin{eqnarray}\label{eq:r}
  r &\equiv& \frac{ \mathcal{P_T}}{ \mathcal{P_R}} =16\omega^2 2^{2(\nu_R-\nu_T)}\frac{\Gamma^2(\nu_T)}{\Gamma^2(\nu_R)}\,,
\end{eqnarray}
where $\mathcal{P}_T(k)$ is the power spectrum of tensor perturbation,
$\Gamma$ is the Gamma function, and $\nu_T$ is given by
\begin{eqnarray}
  \nu_T^2 = a^2H^2\left(2-\omega^2\right)\tau^2+\frac{1}{4}\,.
\end{eqnarray}
Following the same procedure used to demonstrate the near-constancy of
$\dfrac{z''}{z}\,\tau^2$, we apply Eqs.~\eqref{zy3},~\eqref{ea1},
and~\eqref{er5} to the tensor spectral index $\nu_T$. This reveals that $\nu_T$
likewise remains approximately constant near the stable point
$\omega \simeq \lambda / \sqrt{2}$ for $|\lambda|<\sqrt{6}$. Consequently, $r$
can be analytically expressed in terms of $\omega$ and $\lambda$.

We now examine the evolution of the curvature perturbation $\zeta$ on
super-horizon scales, where the $k^2$ term in Eq.~\eqref{eq:evolve_nuk} becomes
negligible. The general solution of $\zeta$ is given by
\begin{eqnarray}
  \zeta_k(\tau) &=& A_k+B_k\int^{\tau}\frac{1}{z^2(\bar{\tau})}d\bar{\tau}\,,
\end{eqnarray}
with integration constants $A_k$ and $B_k$. The conservation of $\zeta$ depends
critically on whether the integral term converges to a constant value as the
system approaches the stable point. Since the integral is monotonically
increasing, if it converges to a constant, this constant can be absorbed into
$A_k$, yielding a decaying mode~(a negative, increasing function). Otherwise,
the integral corresponds a growing mode. Changing the integral variable to
$\omega$, the integral becomes
\begin{eqnarray}
  \int^{\tau}\frac{1}{z^2(\tau)}d\tau &=& \int^{\omega}\frac{1}{z^2}\frac{d\tau}{d\omega}d\omega = \frac{1}{2M_{\rm pl}^2}\int^{t}\frac{1}{a^3\omega^2}dt = \frac{1}{2M_{\rm pl}^2}\int^{\omega}\frac{1}{a^3\omega^2}\frac{1}{\frac{d\omega}{d\phi}\dot{\phi}}d\omega\,.
\end{eqnarray}
Using Eq.~\eqref{ea1}, one can find
\begin{eqnarray}
  \int^{\tau}\frac{1}{z^2(\tau)}d\tau &\sim& \int^{\omega}\frac{1}{\omega^2\left(3-\omega^2\right)}d\omega\,.
\end{eqnarray}
From earlier stability discussion, the integral diverges at the stable point
$\omega = 0$ and $\omega = \pm \sqrt{3}$, while it converges at
$\omega = \frac{\lambda}{\sqrt{2}}$. Therefore, the only physical solution is
the mode evolving toward $\omega=\frac{\lambda}{\sqrt{2}}$, in which the
curvature perturbation is conserved on super-horizon scales.

\begin{figure}[t!]
\centering
\includegraphics[width=3.0in]{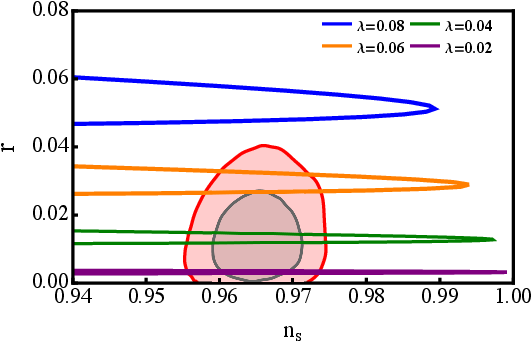}
\includegraphics[width=3.0in]{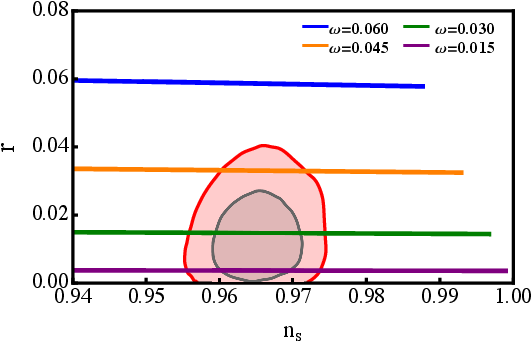}
\caption{Parameter space consistent with the current constraints on $n_{\rm s}$
  and $r$, as inferred from Planck 2018 data~\cite{Planck2018} and a
  combination of BICEP/Keck 2018 and Planck PR4
  data~\cite{Tristram:2021tvh, Tristram:2020wbi, BICEP:2021xfz}. The solid
  contours indicate the 68\% (1$\sigma$) and 95\% (2$\sigma$) confidence
  levels.
}\label{tu1}
\end{figure}

Using Eqs.~\eqref{eq:ns} and~\eqref{eq:r}, the scalar spectral index and the
tensor-to-scalar ratio for different values of $\omega$ and $\lambda$ are
calculated. Figure~\ref{tu1} compares these calculations with the most recent
observational constraints from Planck 2018
data~($n_{\mathrm{s}} = 0.9649 \pm 0.0042$~\cite{Planck2018}) and a combination
of BICEP/Keck 2018 and Planck PR4
data~($r < 0.032$ at 95\% C.L.~\cite{Tristram:2021tvh, Tristram:2020wbi, BICEP:2021xfz}).
This demonstrates that there exists a parameter space that satisfies the
observational constraints. The apparent double-valued behavior for $C>0$ arises
because $n_{\rm s}$ is a complicated, non-monotonic function of $\omega$.
Different values of $\omega$ may correspond to the same $n_{\rm s}$. Each
branch corresponds to a different $\omega$, and within the current theoretical
framework there is no strong reason to prefer one branch over the other. The
choice of branch would ultimately depend on more precise future observational
constraints.

The observational constraints on $n_{\rm s}$ and $r$ impose tight restrictions
on the model parameters, often requiring high-precision tuning. For example, a
finely tuned parameter set
$(\lambda, \omega) = (0.040\pm0.001,\, 0.027\pm0.001)$ yields
$(n_{\rm s}, r) = (0.9700\pm 0.0064,\, 0.012\pm0.001)$, where the uncertainty
of $n_{\mathrm{s}}$ is already larger than that of experimental measurements.
Around this parameter space, we derive an approximate expression for the number
of e-foldings, and subsequently demonstrate that $\omega$ represents an
attractor solution, although it has been implicitly defined in Eq.~\eqref{zd2}.
Assuming the stable point is $\omega = \frac{\lambda}{\sqrt{2}}$, the range of
values for $w$ during inflation is between $[0.027,\,\frac{0.040}{\sqrt{2}}]$,
which leads to
\begin{eqnarray}
\frac{\omega_e}{\omega} \sim \mathcal{O}(1), \quad \frac{\sqrt{3} \pm \omega_e}{\sqrt{3} \pm \omega} \sim \mathcal{O}(1).
\end{eqnarray}
Consequently, the e-foldings in Eq.~\eqref{er2} is approximated to be
\begin{eqnarray}
N_e-N \simeq \frac{2}{6-\lambda^2}\ln\left|\frac{\omega-\frac{\lambda}{\sqrt{2}}}{\omega_e-\frac{\lambda}{\sqrt{2}}}\right|.
\end{eqnarray}

In typical studies, the behavior of $\delta H$ is examined to determine whether
the solution is an attractor, but this approach is not convenient in this work.
The $\delta w$ is instead analyzed. According to Eq.~\eqref{zy3}, $\delta w$
decreases synchronously with $\delta H$, making $\omega$ an equally valid
indicator of attractor behavior.

From the equation of motion, Eq.~\eqref{zd2}, one can derive the differential
relation for $\omega$:
\begin{eqnarray}\label{ed2}
  \frac{d\delta\omega}{\delta\omega} &=& \left(\frac{1}{\omega-\frac{\lambda}{\sqrt{2}}}-\frac{2\omega}{3-\omega^2}-\frac{1}{\omega}\right)d\omega\,,
\end{eqnarray}
Solving the above equation yields the evolution of $\omega$
\begin{eqnarray}
  \delta\omega_e &=& \delta\omega \frac{3-\omega_e^2}{3-\omega^2}\left|\frac{\omega}{\omega_e}\right|\times\left|\frac{\omega-\frac{\lambda}{\sqrt{2}}}{\omega_e-\frac{\lambda}{\sqrt{2}}}\right| \nonumber\\
  &\simeq& \delta\omega\exp\left[-\frac{6-\lambda^2}{2}(N_e-N)\right]\simeq \delta\omega\exp\left[-3(N_e-N)\right]
\end{eqnarray}
Based on the results, the evolution of $w$, which satisfies the constraints on
$n_{\mathrm{s}}$ and $r$, represents an attractor solution. Specifically, $w$
asymptotically approaches the attractor point
$\omega = \frac{\lambda}{\sqrt{2}}$. This reveals that both the general
solutions presented in our work and the particular
solutions~($\omega = \frac{\lambda}{\sqrt{2}}$) discussed in previous studies
belong to the same class of attractors. It is important to emphasize that the
attractor solution in this case should be understood as a family of solutions
rather than a single, isolated trajectory: as the independent variable evolves,
different solutions within this family approach each other
asymptotically~\cite{Liddle:1994dx}.

In summary, despite its simplicity and elegance, power-law inflation was
considered excluded due to the failure of its particular solution to match
observations. We have derived the general solutions of power-law inflation
analytically and demonstrated that power-law inflation is still valid.

The particular solution corresponds to the trajectory where $\omega$ always
equal to $\frac{\lambda}{\sqrt{2}}$, while the general solutions corresponds
to the different trajectories approaching $\omega=\frac{\lambda}{\sqrt{2}}$.
Since the scalar spectral index and tensor-to-scalar ratio are not only
dependent on the attractor itself, but also highly sensitive to the
trajectories how to approach the stable point. It is logically flawed to
exclude a potential model solely based on one trajectory of the attractor
solution that does not meet these constraints. We have proved some of the
trajectories from the general solutions of power-law inflation can satisfy
the observational constraints.

Moreover, power-law inflation can also be analyzed using the slow-roll
approximation~\cite{Liddle:2000cg}. These approximate results are consistent
with those obtained from the exact particular solution~\cite{Weinberg:2008zzc},
but both approaches do not agree with experimental observations. Our work
demonstrates that the slow-roll approximation may fail in certain cases.
Similar issues have been studied in the context of ultra slow-roll
models~\cite{Dimopoulos:2017ged}, where the slow-roll approximation may
also break down.

Finally, our model provides an example showing that certain inflationary
scenarios, previously ruled out under the slow-roll approximation, may in fact
possess viable parameter space with more accurate calculations.

\section*{ACKNOWLEDGMENTS}
The authors gratefully acknowledge Dr. Wei Cheng and Dr. Ruiyu Zhou for their insightful discussions and valuable feedback. Yao Yu,Wen-Zhang Feng and Hong-Song Xie were supported in part by NSFC under Contracts No.~11905023, No.~12047564 and No.~12147102, the Natural Science Foundation of Chongqing (CQCSTC) under Contract No.~cstc2020jcyj-msxmX0555, and the Science and Technology Research Program of Chongqing Municipal Education Commission (STRPCMEC) under Contracts No.~KJQN202200605 and No.~KJQN202200621; Han Zhang and Bai-Cian Ke were supported in part by National Natural Science Foundation of China~(NSFC) under Contracts No.~12192263, Joint Large-Scale Scientific Facility Fund of the NSFC and the Chinese Academy of Sciences under Contract No.~U2032104, and the Excellent Youth Foundation of Henan Scientific Commitee under Contract No.~242300421044.

\end{document}